\documentclass[aps, prx, twocolumn, showpacs]{revtex4-1}
\usepackage{graphicx}
\usepackage{amsmath}
\usepackage{amssymb}
\usepackage{physics}
\usepackage{hyperref}
\usepackage{float}
\usepackage{lipsum}
\hypersetup{colorlinks=true,
	    final=true,
	    linkcolor=blue,
	    citecolor=blue,
	    filecolor=blue,
	    urlcolor=blue,
}

\begin{document}
\title{Many-body electronic structure of $d^{9-\delta}$ layered nickelates}
\author{Harrison LaBollita}
\email{hlabolli@asu.edu}
\affiliation{Department of Physics, Arizona State University, Tempe, AZ 85287, USA}
\author{Myung-Chul Jung}
\affiliation{Department of Physics, Arizona State University, Tempe, AZ 85287, USA}
\author{Antia S. Botana}
\affiliation{Department of Physics, Arizona State University, Tempe, AZ 85287, USA}
\date{\today}

\begin{abstract}
The recent observation of superconductivity in an infinite-layer and quintuple-layer nickelate within the same $R_{n+1}$Ni$_{n}$O$_{2n+2}$ series ($R$ = rare-earth, $n=2-\infty$, with $n$ indicating the number of NiO$_{2}$ layers along the $c$-axis), unlocks their potential to embody a whole family of unconventional superconductors. Here, we systematically investigate the many-body electronic structure of the layered nickelates (with $n=2-6,\infty$) within a density-functional theory plus dynamical mean-field theory framework and contrast it with that of the known superconducting members of the series and with the cuprates. We find that many features of the electronic structure are common to the entire nickelate series, namely, strongly correlated Ni-$d_{x^{2}-y^{2}}$ orbitals that dominate the low-energy physics, mixed Mott-Hubbard/charge-transfer characteristics, and $R$($5d$) orbitals acting as charge reservoirs. Interestingly, we uncover that the electronic structure of the layered nickelates is highly tunable as the dimensionality changes from  quasi-two-dimensional to three-dimensional as $n \rightarrow \infty$. Specifically, we identify the tunable electronic features to be: the charge-transfer energy, presence of $R(5d)$ states around the Fermi level, and the strength of electronic correlations.

\end{abstract}
\maketitle

\section{\label{sec:intro}Introduction}
The search for materials with analogous structural, magnetic, and electronic motifs to the high-$T_{c}$ cuprates has been an active field of research for more than 30 years \cite{Bednorz1986,Keimer2015}. Nickel oxide (nickelate) materials have been an obvious class of candidate materials as nickel sits next to copper in the periodic table, and it can realize a Ni$^{+}$: $d^9$ oxidation state, which is isoelectronic with Cu$^{2+}$ \cite{anisimov1999,pickett2004}.

The promise of nickelates in this context was realized in 2019 with the observation of superconductivity in Sr-doped NdNiO$_{2}$, referred to as the `infinite-layer' ($n= \infty$) nickelate \cite{Li2019}, with a Ni$^{+}$ ($d^9$) oxidation state and NiO$_2$ planes akin to the CuO$_2$ planes of the cuprates. Superconductivity in the infinite-layer material displays a dome-like dependence with a maximum $T_{c}$ near 20\% hole-doping or a $d^{8.8}$ electron filling \cite{Li2020dome}. Since this initial discovery, superconductivity has been observed in other rare-earth flavors of the infinite-layer nickelate, namely hole-doped PrNiO$_{2}$ \cite{Osada2020} and LaNiO$_{2}$ \cite{Osada2021nickelate,Zeng2021superconductivity}, all with similar $T_{c}$'s and superconducting domes. 
\begin{figure}
    \centering
    \includegraphics[width=\columnwidth]{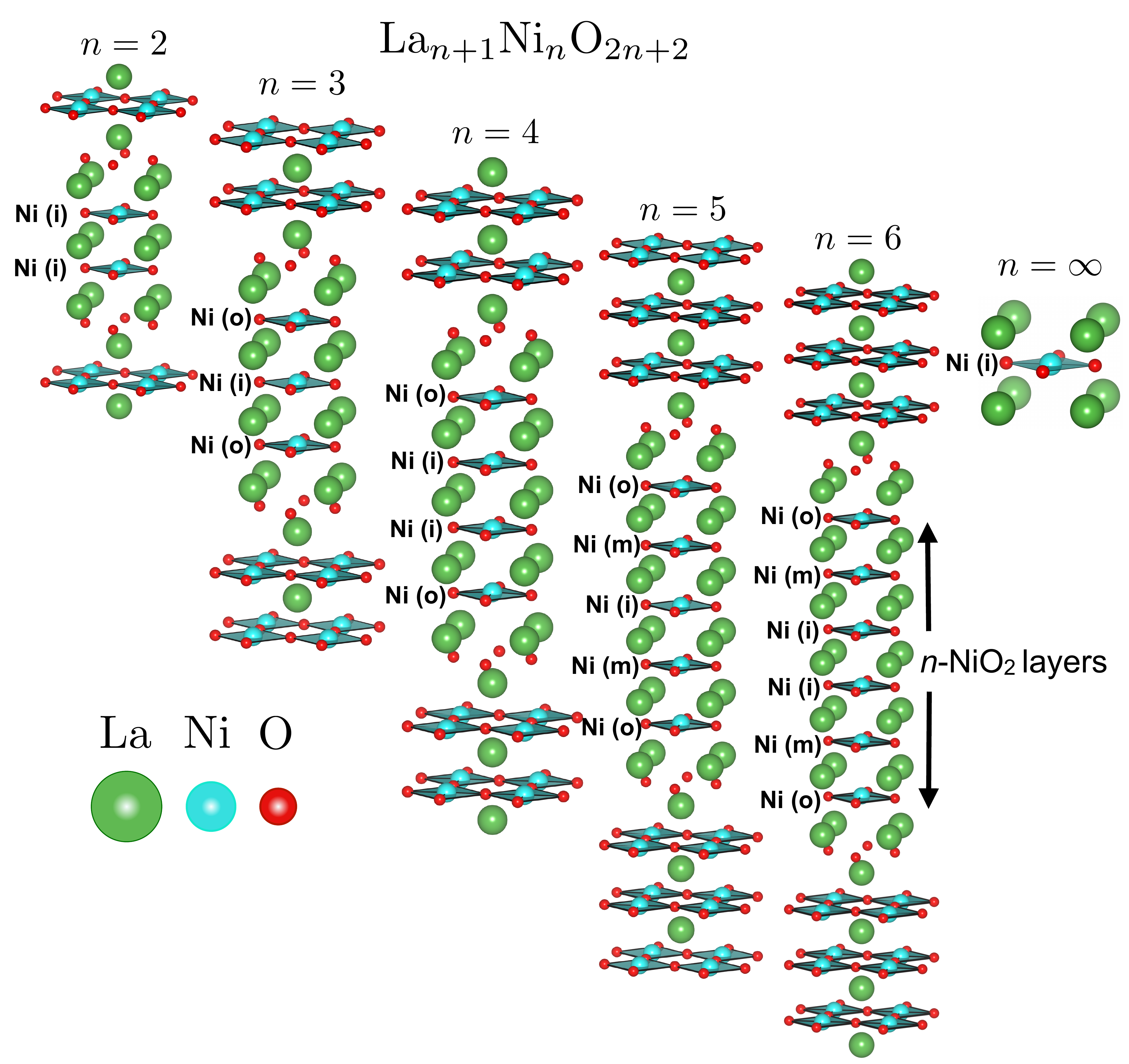}
    \caption{Crystal structures of the layered nickelates La$_{n+1}$Ni$_{n}$O$_{2n+2}$ for $n=2-6,\infty$ (from left to right). Each crystal structure contains $n$ NiO$_{2}$ planes interleaved between R-O fluorite slabs. The stoichiometries for the nickelate series are: La$_{3}$Ni$_{2}$O$_{6}$  ($n=2$), La$_{4}$Ni$_{3}$O$_{8}$  ($n=3$), La$_{5}$Ni$_{4}$O$_{10}$ ($n=4$), La$_{6}$Ni$_{5}$O$_{12}$ ($n=5$), La$_{7}$Ni$_{6}$O$_{14}$ ($n=6$), and LaNiO$_{2}$ ($n=\infty$). Note for even $n$ there is no mirror plane. Ni(i), Ni(m), and Ni(o) denote the inner, middle, and outer inequivalent Ni sites, respectively.}
    \label{fig:struct}
\end{figure}
The discovery of superconductivity in the infinite-layer nickelates has attracted a great deal of theoretical \cite{wu2019, Hu2019, jiang2019, nomura2019,  Choi2020, Ryee2020, Gu2020, Karp2020_112, botana2020, Leonov2020, Kapeghian2020, lechermann2020late, lechermann2020multi, pickett2020, kitatani2020, petocchi2020,zhang2020self, Sakakibara2020, jiang2020, werner2020, Zhang2020eff, Zhang2020, Wang2020, Bandyopadhyay2020, Been2021} and experimental attention \cite{Li2019, Fu2020, Osada2020, lee2020aspects, goodge2020, hepting2020, Fu2020, Li2020, Li2020dome, BiXiaWang2020, QiangqiangGu2020, cui2020nmr, Liu2020, Osada2021nickelate, Zeng2021superconductivity}. Importantly, the infinite-layer nickelate belongs to a large series of layered nickelate materials with the general chemical formula, $R_{n+1}$Ni$_{n}$O$_{2n+2}$ ($R$ = La, Pr, Nd; $n=2,3,\dots,\infty$). The finite-layer ($n \neq \infty$) nickelates have $n$-quasi-2D NiO$_{2}$ planes (like the $n=\infty$ material) but display an additional blocking $R$-O fluorite slab interleaved between neighboring blocks of NiO$_{2}$ planes (see Fig. \ref{fig:struct}). The $R$-O layer tunes the dimensionality of the materials as $n$ decreases: from three-dimensional-like in the $n = \infty$ compound to quasi-two-dimensional as $n$ decreases. In addition, the number of NiO$_{2}$ planes along the $c$-axis ($n$) tunes the average nominal Ni filling that can be defined as $d^{9-\delta}$, where $\delta=1/n$. Notably, the quintuple-layer $(n=5)$ nickelate has a nominal $d^{8.8}$ ($\delta=0.2$) Ni($3d$) filling that matches the optimal doping level of the infinite-layer nickelates and the cuprates. Indeed, recently, superconductivity has been observed in the quintuple-layer nickelate Nd$_{6}$Ni$_{5}$O$_{12}$ without the need for chemical doping \cite{pan2021super}. The fact that superconductivity has now been demonstrated in two members of the layered nickelate series (the only ones so far where an optimal $d^{8.8}$ filling has been attained) suggests that a new family of superconductors has been uncovered.

In this paper, we systematically investigate the many-body electronic structure of the layered nickelates ($n=2-6,\infty$) by including electronic correlations beyond density-functional theory (DFT) within a DFT plus dynamical mean-field theory (DMFT) framework and contrast it with that of the known superconducting members of the series and with the cuprates. We find that there are many electronic features that span across the entire nickelate series, namely, strongly correlated Ni-$d_{x^{2}-y^{2}}$ orbitals that dominate the low-energy physics similar to the Cu-$d_{x^{2}-y^{2}}$ orbitals of the cuprates, as well as mixed Mott-Hubbard/charge-transfer characteristics and La($5d$) orbitals acting as charge reservoirs. We uncover certain features of the electronic structure of layered nickelates are tunable with $n$, specifically, the charge-transfer energy, involvement of $R(5d)$ states around the Fermi level, and the strength of electronic correlations. All in all, with the dimensionality (via the number of layers, $n$) controlling the electronic structure of this nickel oxide family, we anticipate that other layered nickelates should be able to host a superconducting instability provided that the proper Ni electron filling (near $d^{8.8}$) can be achieved. 

\section{\label{sec:method}Methodology}
The charge self-consistent (CSC) combination of density-functional theory plus dynamical mean-field theory (DFT+DMFT) is employed to calculate the many-body electronic structure of the layered nickelates ($n=2-6,\infty$). The DFT problem is solved using the all-electron, full-potential code {\sc wien}2k \cite{wien2k}, which is built on the augmented plane wave plus local orbital basis set (APW+lo). The Perdew-Burke-Ernzerhof (PBE) version \cite{pbe} of the generalized gradient approximation (GGA) is used as the exchange-correlation functional. We choose $R$ = La to avoid any ambiguity in the treatment of $4f$ electrons that would arise from $R$ = Nd or Pr. It seems reasonable to choose $R$ = La for a general description of the many-body electronic structure, as the electronic structure of the infinite-layer compounds has been shown not to significantly change with $R$ \cite{Kapeghian2020,Bandyopadhyay2020,Been2021}.
A dense $k$-mesh of 17$\times$17$\times$17 is used for integration in the Brillioun zone. We used $R_{\mathrm{MT}}K_{\mathrm{max}} = 7$ and muffin-tin radii of 2.35, 1.97, and 1.75 a.u. for La, Ni, and O, respectively.

To gain quantitative insights into the electronic structure of the layered nickelates, we construct maximally localized Wannier functions (MLWFs) from the DFT spectrum.  We employed {\sc wannier}90 \cite{wannier90} and {\sc wien}2{\sc wannier} \cite{wien2wannier} to obtain the MLWFs. We compute the MLWFs within a wide energy window, including the Ni($3d$), O($2p$), and La($5d)$ orbitals, which provides orbital energies and allows us to estimate the charge-transfer energy. We obtained well-localized (albeit not unique) Wannier functions that correctly reproduce the DFT band structure. %For tight-binding hopping parameters, we use only the Ni-$d_{x^2-y^2}$ orbital(s) for the initial projection within the smaller energy window.

For the DMFT part, we construct a basis of atomic-like orbitals from the Kohn-Sham wavefunctions using a projector scheme \cite{triqs_dft_tools, triqs_wien2k_interface} spanning a correlated subspace of size $-10$ eV to $10$ eV around the Fermi level. Each inequivalent Ni site in the crystal structure is treated as its own quantum impurity problem, where the full Ni($3d$) manifold is treated as correlated and governed by a fully rotationally-invariant Slater Hamiltonian. We parameterize the Slater integrals with parameters representative of the nickelates: a Hubbard $U = F^{0} = 7$ eV and Hund's coupling $J_{\mathrm{H}} = (F^{2} + F^{4})/14 = 0.7$ eV \cite{nowadnick2015,Karp2020_438,Karp2020_112,Karp2021}. The quantum impurity problem(s) are solved using a continuous-time quantum Monte Carlo (QMC) algorithm based on the hybridization expansion method as implemented in TRIQS/{\sc cthyb} \cite{triqs, cthyb}. To reduce high-frequency noise in the QMC data, we represent both the Green's function and self-energies in a basis of Legendre polynomials and sample the Legendre coefficients directly within the TRIQS/{\sc cthyb} solver \cite{Boehnke2011legendre}. The fully-localized limit (FLL) formula is used for the double counting correction. All calculations are performed at a system temperature of 290 K ($\beta = 40$ eV$^{-1}$) in the paramagnetic state. Maximum entropy methods are used to analytically continue the QMC data from Matsubara space to real-frequency space  \cite{triqs_maxent}.

\begin{figure*}
\centering
\includegraphics[width=2\columnwidth]{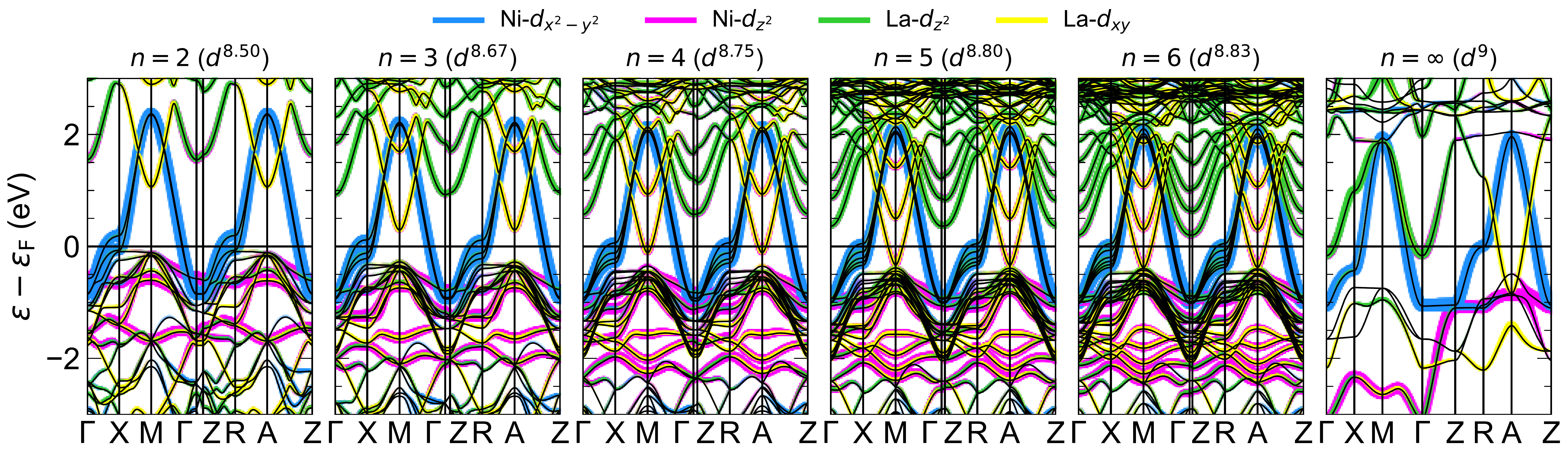}
\caption{DFT band structures for the $n=2-6,\infty$ (left to right) layered nickelates within the `fatband' representation, which highlights the orbital character of the DFT bands. The orbital character of the Ni-$d_{x^{2}-y^{2}}$ (blue), Ni-$d_{z^{2}}$ (pink), La-$d_{z^{2}}$ (green), and La-$d_{xy}$ (yellow) orbitals are shown.}
\label{fig:dft}
\end{figure*}

\section{Crystal structure and ionic count}
The layered nickelates are derived from either a perovskite ($n=\infty$) or Ruddlesden-Popper ($n \neq \infty$) parent compound via oxygen deintercalation. Currently, the $n=2-5$ parent Ruddlesden-Popper phases of the series have been synthesized in thin film geometry for $R=$ La and Nd \cite{Nie2020,Pan2022}. These reduced $R_{n+1}$Ni$_{n}$O$_{2n+2}$ compounds crystallize in a tetragonal structure belonging to either the I4/mmm ($n \neq \infty$) or P4/mmm ($n=\infty$) space group. As mentioned above, the series of layered nickelates contains blocks of $n$ two-dimensional NiO$_{2}$ planes (with square-planar coordination) separated by rare-earth layers. For the finite-layer nickelates, an extra rare-earth fluorite slab is present, separating each block of $n$ $R$-NiO$_{2}$ planes (see Fig. \ref{fig:struct}). Furthermore, each $R$-NiO$_{2}$ block is shifted by half a lattice constant in the $ab$-plane. These two additional structural features  (that are not present in the infinite-layer nickelate)  effectively decouple neighboring $R$-NiO$_{2}$ blocks along the $c$-direction. 

As mentioned above, the average nominal occupation on the Ni sites for each layered nickelate is $d^{9-\delta}$, where $\delta = 1/n$. Specifically, the average oxidation states for $n= 2-6,\infty$ layered nickelates are: Ni$^{1.5+}$ ($d^{8.5}$) for $n=2$ (La$_{3}$Ni$_{2}$O$_{6}$), Ni$^{1.33+}$ ($d^{8.67}$) for $n=3$ (La$_{4}$Ni$_{3}$O$_{8}$), Ni$^{1.25+}$ ($d^{8.75}$) for $n=4$ (La$_{5}$Ni$_{4}$O$_{10}$), Ni$^{1.2+}$ ($d^{8.8}$) for $n=5$ (La$_{6}$Ni$_{5}$O$_{12}$), Ni$^{1.17+}$ ($d^{8.83}$) for $n=6$ (La$_{7}$Ni$_{6}$O$_{14}$), and Ni$^{+}$ ($d^{9}$) for $n=\infty$ (LaNiO$_{2}$).

Each material contains either an inner (i), inner (i) and outer (o), or an inner (i), middle (m), and outer (o) symmetry-inequivalent Ni site along the $c$-axis. For odd $n$ materials, the inner Ni site acts as a mirror plane for the outer layer(s) (see Fig. \ref{fig:struct}). We use the experimental crystal structures for the $n=2,3,\infty$ layered nickelates \cite{poltavets2006,poltavets2007,hayward1999}. As the structures of lanthanum-based $n=4-6$ layered nickelates have not been experimentally resolved yet, we derive their crystal structures from the experimental structure of the $n=3$ material La$_{4}$Ni$_{3}$O$_{8}$.  For further details on the derivation of the crystal structures of the $n=4-6$ materials see Ref. \onlinecite{labollita2021}.

\section{\label{sec:dft}non-interacting electronic structure}
Figure \ref{fig:dft} summarizes the electronic structure of the $n=2-6, \infty$ layered nickelates at the DFT level. Starting with the $n=2$ material with an average Ni($3d$) occupation of $d^{8.50}$ ($\delta = 0.5$), only two partially filled $d_{x^{2}-y^{2}}$ bands (one per Ni) cross the Fermi level, in a cuprate-like fashion. The splitting between these bands (highlighted at X) is a consequence of the interlayer hopping within NiO$_{2}$ blocks and is analogous to that of the multi-layer cuprates \cite{Sakakibara2014}. 
Below $\varepsilon_{\mathrm{F}}$, there is a complex of mostly filled Ni-$t_{2g}$ bands and  particularly flat Ni-$d_{z^{2}}$ bands highlighted in pink. The flat dispersion of the Ni-$d_{z^{2}}$ is a unique feature of the finite-layer nickelates, dissimilar to the infinite-layer compound, which has a dispersive band coming from the Ni-$d_{z^{2}}$ orbital along the $\Gamma-$Z direction. The flatness of these bands is a consequence of the fluorite slab, present in all finite-layer nickelates, which cuts the $c$-axis dispersion across the neighboring NiO$_{2}$ blocks, as described above. Above $\varepsilon_{\mathrm{F}}$, there are a set of bands of mostly $d_{xy}$ and $d_{z^{2}}$ orbital character stemming from the La sites within the NiO$_{2}$ blocks. 

As the number of  NiO$_{2}$ layers increases with increasing $n$, we find many of the same features: $n$ partially filled $d_{x^{2}-y^{2}}$ bands coming from the Ni sites, flat Ni-$d_{z^{2}}$ bands, mostly filled Ni-$t_{2g}$ bands, and  La($5d$) bands above $\varepsilon_{\mathrm{F}}$. 
However, a key difference with increasing $n$ is the location of La($5d$) bands as they experience a significant downward shift in energy with increasing $n$. For the $n=4$ material, a band with La-$d_{xy}$ character already crosses $\varepsilon_{\mathrm{F}}$ giving rise to an electron pocket at the zone corners of the Brillouin zone that self-dopes the $d_{x^{2}-y^{2}}$ bands. Moving onto the $n=5$ and 6 materials, these electron pockets continue to increase in size as the bands with mostly La-$d_{xy}$ character shift even deeper in energy. Across the entire $n=2-6,\infty$ series, the $n=2,3$ materials are then markedly different as they lack this self-doping effect. In fact, because the trilayer ($n=3$) nickelate does not have La($5d$) bands crossing the Fermi level, and given its close in proximity to the cuprate superconducting dome in terms of Ni($3d$) electron count, previous works have suggested this material to be one of the closest cuprate analogs to date \cite{botana2017,Zhang2017,dean2021,dean2022}. Furthermore, as $n$ reaches the infinite-layer limit the electronic structure changes from quasi-two-dimensional to three-dimensional \cite{labollita2021}. This can be easily observed by looking at the bands around the Fermi level in the $k_{z}=0$ ($\Gamma$-X-M-$\Gamma$) and $k_{z}=1/2$ (Z-R-A-Z) planes for the infinite-layer material. The bands in these two $k_{z}$ planes would generate distinctly different constant energy surfaces and therefore a rather three-dimensional-like electronic structure relative to the $n\neq\infty$ nickelates.

\begin{figure*}
    \centering
    \includegraphics[width=2\columnwidth]{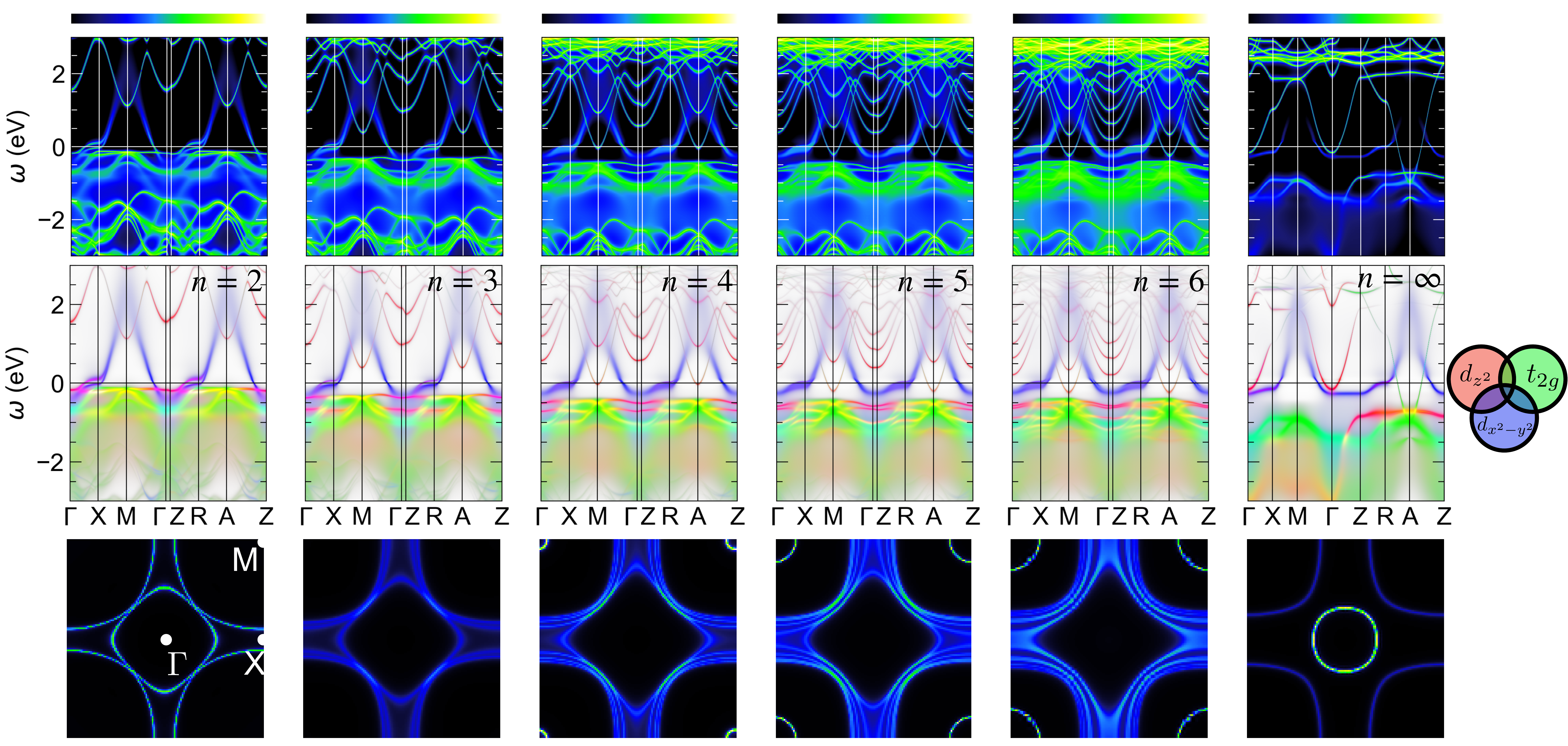}
    \caption{Top panels: ${\bf k}$-resolved spectral functions, $A({\bf k}, \omega)$, along high-symmetry lines in the Brillioun zone for $n=2-6,\infty$ layered nickelates (from left to right). Middle panels: Orbital-projected $A({\bf k},\omega)$ for the Ni($3d$) shell with: blue denoting the $d_{x^{2}-y^{2}}$ orbitals, red denoting the $d_{z^{2}}$ orbitals, and green denoting the $t_{2g}$ orbitals. Bottom panels: Corresponding interacting Fermi surfaces $A({\bf k}, \omega=0)$ in the $k_{z} = 0$ plane. The high-symmetry points in the BZ are denoted.}
    \label{fig:Akw}
\end{figure*}

According to the Zaanen-Sawatsky-Allen (ZSA) classification scheme \cite{allen1985}, cuprates sit well within the charge-transfer regime ($U \gg \Delta_{\mathrm{CT}}$) with the on-site Coulomb repulsion $U$ being larger than the the charge-transfer energy  ($\Delta_{\mathrm{CT}}$) that measures the hybridization between the TM($d$) and ligand($p$) orbitals. In this regard, the hybridization of the O($2p$) orbitals with the Cu($3d$) orbitals is believed to be an essential ingredient for the formation of Zhang-Rice (ZR) singlets and for cuprate superconductivity. At the level of DFT, the hybridization between the Ni($3d$) and O($2p$) orbitals can be quantified by $\Delta_{\mathrm{CT}} = \varepsilon_{d} - \varepsilon_{p}$, where $\varepsilon_{d(p)}$ refers to the on-site orbital energy for the Ni-$d_{x^{2}-y^{2}}$ (O-$p_{\sigma}$) orbital derived from fitting maximally localized Wannier functions (MLWFs) to the DFT spectrum. %For more details on the construction of MLWFs see Appendix \ref{sec:app_A}.
For cuprates, a typical value of $\Delta_{\mathrm{CT}}$ is $1 - 2$ eV. For the infinite-layer nickelate, values of $4.4 - 5$ eV have been obtained with similar methods \cite{lechermann2020late, nica2020}. We note that the larger $p-d$ splitting in the infinite-layer nickelates has been identified as one of the main differences with the cuprates \cite{nica2020,pickett2004,dean2022,goodge2020,hepting2020}.
Here, we obtain charge-transfer energies of $\Delta_{\mathrm{CT}}(n=2) = 3.3$ eV, $\Delta_{\mathrm{CT}}(n=3) = 3.7$ eV, $\Delta_{\mathrm{CT}}(n=4) = 3.8$ eV, $\Delta_{\mathrm{CT}}(n=5) = 3.9$ eV, $\Delta_{\mathrm{CT}}(n=6) = 4.0$ eV, and $\Delta_{\mathrm{CT}}(n=\infty) = 4.5$. As such, as $n$ increases from 2 to $\infty$, the charge-transfer energy increases by $\sim 1$ eV indicating the increasing hybridization between the Ni($3d$) and O($2p$) orbitals for smaller $n$ (larger $\delta$) nickelates i.e. when going towards the 2D limit. This trend in charge-transfer energies (increasing with $n$) qualitatively agrees with the available experimental data for the $n=3$, $n=5$, and $n=\infty$ materials \cite{pan2021super}. Assuming a reasonable Coulomb repulsion of $U = 5-7$ eV \cite{nowadnick2015}, this places the layered nickelates at the boundary between the charge-transfer and Mott-Hubbard regimes in the ZSA scheme ($U \sim \Delta$). Interestingly, even though these energies are larger than in the cuprates, they still lead to sizeable superexchanges \cite{dean2021,Lu2021}. Therefore, this suggests that the O($2p$) states remain an appreciable ingredient for the electronic structure of the layered nickelates \cite{Karp2020_438, dean2022}.

\section{\label{sec:dmft}DFT+DMFT results}
\subsection{Spectral properties and electronic correlations}
\begin{figure}
    \centering
    \includegraphics[width=\columnwidth]{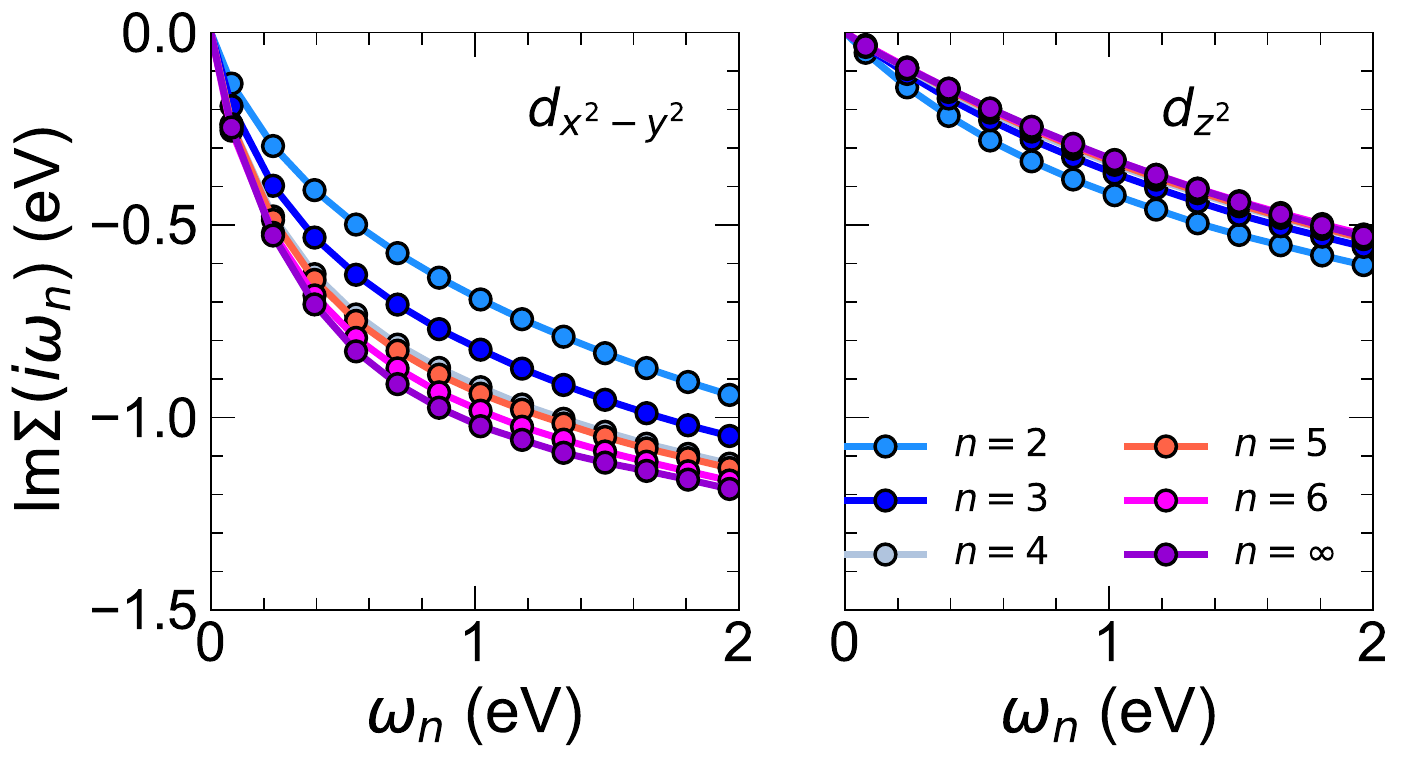}
    \caption{Imaginary part of the Ni self-energies (averaged over inequivalent Ni sites) for the $e_{g}$ manifold in Matsubara space: $d_{x^{2}-y^{2}}$ (left) and $d_{z^{2}}$ (right).}
    \label{fig:mstar}
\end{figure}

With the non-interacting electronic structure of the layered nickelates established, we now analyze the interacting problem to investigate the role of electronic correlations. Figure \ref{fig:Akw} summarizes the ${\bf k}$-resolved spectral functions, $A({\bf k}, \omega)$ for the $n=2-6,\infty$ layered nickelates. The low-energy spectrum exhibits coherent quasiparticle (QP) dispersions corresponding to the Ni-$d_{x^{2}-y^{2}}$ states. Compared to the DFT spectrum, the Ni-$d_{x^{2}-y^{2}}$ states are strongly renormalized. Mass enhancements are derived from the electronic self-energies, $m^{\star}/m_{\mathrm{DFT}} = 1 - \partial\mathrm{Im}\Sigma(i\omega_{n}\rightarrow 0)/\partial \omega_{n}$ shown in Fig. \ref{fig:mstar}. We find the mass enhancement for the Ni-$d_{x^2-y^2}$ orbital monotonically increases from $m^{\star}/m_{\mathrm{DFT}} = 2.5$ to $3.7$ as $n$ increases from 2 to $\infty$, which is in agreement with previous DFT+DMFT calculations \cite{Karp2020_112, Karp2020_438, Karp2021, labollita2022}. Note that both the $e_{g}$ self-energies and mass enhancements have been averaged over inequivalent Ni impurities. Thus, the strength of electronic correlations on the Ni-$d_{x^{2}-y^{2}}$ states increases with $n$, which is the expected trend for a mixed Mott-Hubbard/charge-transfer system as we approach the $d^{9}$ ($\delta=0$) limit. This monotonic increase in mass enhancements with $n$ matches previous works studying the hole-doping effects on the electronic structure of infinite-layer nickelates \cite{Leonov2020,Wang2020}. The mass enhancements on all other Ni($3d$) orbitals are much smaller and remain $\sim 1.5-1.7$. 

Figure \ref{fig:Akw} (middle panels) also shows the orbital character of the spectral weight using the `fatspec' representation introduced in Ref. \onlinecite{lechermann2022}, where blue corresponds to the Ni-$d_{x^{2}-y^{2}}$ orbitals, red to the Ni-$d_{z^{2}}$ orbitals, and green to the Ni-$t_{2g}$ orbitals. For the finite-layer nickelates, apart from the Ni-$d_{x^{2}-y^{2}}$ states dominating around the Fermi level, the spectral functions reveal non-dispersing spectral weight corresponding to the Ni-$d_{z^{2}}$ states located just below these metallic Ni-$d_{x^{2}-y^{2}}$ states. This flat Ni-$d_{z^{2}}$ spectral weight retreats away from the chemical potential as $n$ increases. Similarly, the spectral weight corresponding to the Ni-$t_{2g}$ states experiences a downward shift in the energy spectrum as $n$ is increased. For the infinite-layer material, the spectral weight corresponding to the Ni-$d_{z^{2}}$ orbital is dispersive along the $k_{z}$ direction and becomes flat only in the $k_{z}=1/2$ plane.

\begin{figure*}
    \centering
    \includegraphics[width=1.75\columnwidth]{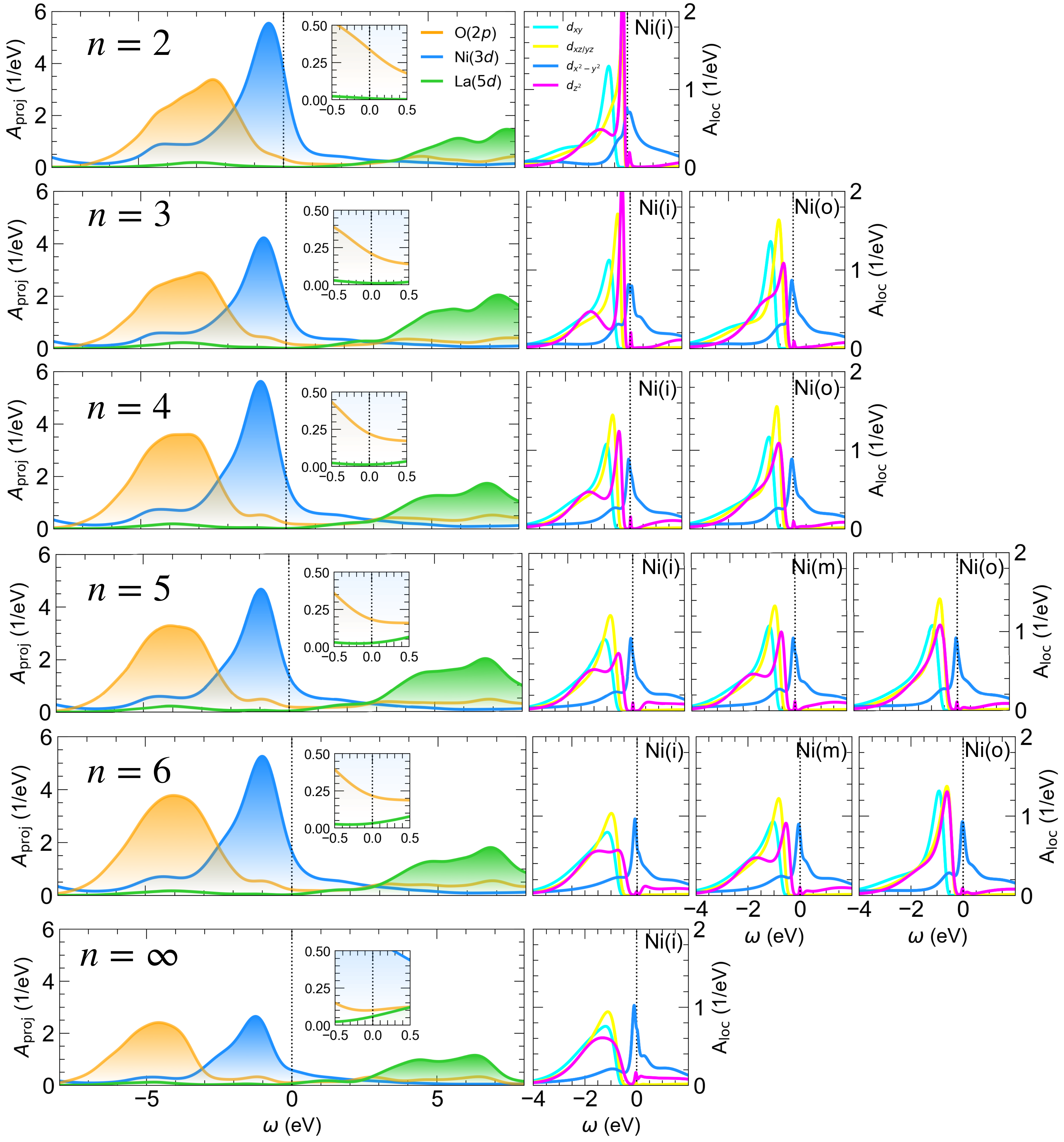}
    \caption{Left panels: Orbital-resolved {\bf k}-integrated spectral functions for the $n=2-6,\infty$ layered nickelates. The inset shows the orbital-projected spectral weight corresponding to the O($2p$) and La($5d$) states around the Fermi level. Right panels: Impurity-resolved local spectral function(s) for the Ni($3d$) manifold.}
    \label{fig:Aw}
\end{figure*}

The impurity-resolved local spectral functions shown in Fig. \ref{fig:Aw} offer further insights into the low-energy physics of the layered nickelates. The component corresponding to the Ni-$d_{x^{2}-y^{2}}$ orbital is dominant at low-energy and exhibits the characteristic three-peak structure of a mixed Mott-Hubbard/charge-transfer system. There is some minor Ni-$d_{z^{2}}$ spectral weight but this arises from the small hybridization between the Ni-$d_{x^{2}-y^{2}}$ and Ni-$d_{z^{2}}$ states which can be seen from the fatspec representation of the ${\bf k}$-resolved spectral function in Fig. \ref{fig:Akw}. The local spectral function highlights the evolution of the flat Ni-$d_{z^{2}}$ band captured in Fig. \ref{fig:Akw} stemming from the innermost Ni site. For the $n=2$ material, this flat band is closest to the Fermi level and causes a large peak in the Ni-$d_{z^{2}}$ local spectral function. This peak softens as this band becomes more dispersive with increasing $n$ as the electronic structure becomes more three-dimensional-like. The Ni-$t_{2g}$ states remain mostly filled and inert by only shifting slightly downward in the energy spectrum with increasing $n$.

For the $n=4-6,\infty$ nickelates, the La($5d$) states survive electronic correlations and remain active crossing the chemical potential. Despite the fact that these states exhibit some Ni-$d_{z^{2}}$ weight (see Fig. \ref{fig:Akw}), the Ni-$d_{z^{2}}$ states remain only weakly correlated meaning that a significant correlation-induced shift is not likely to occur. Previous works on the $n=5$ material have shown that these La($5d$) states can be lifted away from the chemical potential by either treating the La($5d$) orbitals as correlated \cite{worm2021correlations} or within a DFT+sicDMFT framework with strong coupling \cite{lechermann2022}. Whether $R(5d)$ states do indeed contribute to the Fermi surface is an open question and can only be settled with further experiments. 

In this context, we calculate the interacting Fermi surface, $A({\bf k}, \omega=0)$ in the $k_{z}=0$ plane for each material as shown in Fig. \ref{fig:Akw} (bottom panels). For all $n$, there are large hole-like sheets (arcs) coming from the Ni-$d_{x^{2}-y^{2}}$ states, whose van Hove singularity lies below the chemical potential and a single electron-like sheet (square-like) coming from the Ni-$d_{x^{2}-y^{2}}$ states whose van Hove singularity sits above the chemical potential. As we increase $n$, the splitting between the Ni-$d_{x^{2}-y^{2}}$ quasiparticle dispersions decreases and the topology of this electron-like sheet transforms as $n$ approaches the infinite-layer limit. Additionally, there are electron pockets at the zone corners of La($5d$) character previously discussed for the $n=4-6$ materials (that increase in size as $n$ increases). For the infinite-layer material, the electron-like pocket at the zone center comes from the La($5d$) band crossing at $\Gamma$ of $d_{z^2}$ character (see Fig. \ref{fig:Akw}). The electron-like pockets from the La($5d$) band crossing at A of $d_{xy}$ character is only visible in the $k_{z}=1/2$ plane, see for example Ref. \onlinecite{Karp2020_438}. Overall, the fermiology matches the Fermi surfaces calculated within DFT \cite{labollita2021}. One important difference between the fermiology of the finite-layer nickelates when compared to the infinite-layer compound is the overall dimensionality. The finite-layer nickelates have a two-dimensional-like fermiology, very similar to that of the multi-layer cuprates, whereas the fermiology of the infinite-layer nickelates is 3D-like \cite{Sakakibara2014}. This reduction in  dimensionality for the finite-layer materials is a consequence of the difference in structure described previously and further strengthens the cuprate-like nature of the finite-layer nickelates \cite{labollita2021}.

\begin{table}
    \centering
    \begin{tabular*}{\columnwidth}{l@{\extracolsep{\fill}}lcccc}
    \hline
    \hline
         material                  & Ni site & $n_{d_{x^{2}-y^{2}}}$ & $n_{d_{z^{2}}}$ & $n_{t_{2g}}$ & $n_{\mathrm{tot}}$  \\
         \hline
         La$_{3}$Ni$_{2}$O$_{6}$   & Ni(i)   & 0.96 & 1.67 & 5.84 & 8.47 \\
         \hline
         La$_{4}$Ni$_{3}$O$_{8}$   & Ni(i)   & 1.00 & 1.62 & 5.82 & 8.44 \\
          --                       & Ni(o)   & 1.03 & 1.66 & 5.83 & 8.52 \\
         \hline
         La$_{5}$Ni$_{4}$O$_{10}$  & Ni(i)   & 1.06 & 1.61 & 5.81 & 8.49 \\
          --                       & Ni(o)   & 1.06 & 1.65 & 5.81 & 8.52 \\
         \hline
          La$_{6}$Ni$_{5}$O$_{12}$ & Ni(i)   & 1.10 & 1.60 & 5.80 & 8.50 \\
          --                       & Ni(m)   & 1.08 & 1.60 & 5.80 & 8.49 \\
          --                       & Ni(o)   & 1.05 & 1.64 & 5.80 & 8.50 \\
          \hline
          La$_{7}$Ni$_{6}$O$_{14}$ & Ni(i)   & 1.11 & 1.60 & 5.80 & 8.51 \\
          --                       & Ni(m)   & 1.09 & 1.61 & 5.81 & 8.51 \\
          --                       & Ni(o)   & 1.06 & 1.65 & 5.81 & 8.52 \\
          \hline
          LaNiO$_{2}$              & Ni(i)   & 1.14 & 1.58 & 5.78 & 8.51 \\
          \hline
          \hline
    \end{tabular*}
    \caption{Impurity- and orbital- resolved Ni($3d$) occupations for the $n=2-6,\infty$ layered nickelates computed from the impurity Green's functions.}
    \label{tab:occupations}
\end{table}

\subsection{The role of La$(5d)$ and O$(2p)$ states}

Changing the number of layers along the $c$-axis ($n$) effectively tunes the formal charge on the Ni sites: as $n$ decreases from $n=\infty$ to $n=2$, one is essentially hole-doping the nickelate system by  $\delta=1/n$, as mentioned above. The consequences of this `hole-doping' effect are revealed in the overall ${\bf k}$-integrated spectral functions (see Fig. \ref{fig:Aw}, left panel). Starting with the bilayer ($n=2$) material, the Ni($3d$) states dominate the low-energy physics. Most of the spectral weight around $\omega=0$ corresponds to the Ni-$d_{x^{2}-y^{2}}$ orbitals, as revealed in the local spectral function (see Fig. \ref{fig:Aw}, right panel). In the addition spectrum ($\omega > 0$), the La($5d$) states are far above the chemical potential. The states corresponding to the Ni-$t_{2g}$ orbitals contribute a large peak just below the chemical potential. The O($2p$) states dominate the remainder of the removal spectrum ($\omega < 0$) exhibiting a broad peak centered around $\omega=-3$ eV. Upon increasing $n$, the complex of O($2p$) states shifts downward in the spectrum, while the La($5d$) states become active around the chemical potential, giving rise to the self-doping effect described above. The shift in O($2p$) states qualitatively matches the increasing charge-transfer energy with $n$  discussed in the previous section. Additionally, the qualitative trends in the electronic spectrum as $n$ changes agree with hole-doping studies of the infinite-layer nickelate, pointing to the importance of making comparisons at the same Ni($3d$) filling \cite{kitatani2020, Karp2020_438, labollita2022}.

While the metallic Ni-$d_{x^{2}-y^{2}}$ states are dominant around the chemical potential, there is some non-zero spectral weight corresponding to the O($2p$) states present at all $n$ (see insets shown in Fig. \ref{fig:Aw}). This spectral feature is reminiscent of the ZR physics of the cuprates and has been observed in spectroscopic measurements of the hole-doped infinite-layer nickelates \cite{goodge2020}. Additionally, there is a small amount of spectral weight corresponding to La($5d$) states, which increases with $n$. For the cuprates, doped holes reside on the oxygen sites and act as a charge-reservoir in addition to providing bandwidth to the Cu($3d$) states. In the case of the nickelates, where both O($2p$) and La($5d$) states are present near the chemical potential, the Ni($3d$) bandwidth is provided through hybridization with the O($2p$) states, analogous to the cuprates and supported by experiment \cite{hepting2020}. However, unlike the cuprates, the La($5d$) states act as a charge reservoir absorbing most of the dopants \cite{Karp2020_438}.

\begin{figure}
    \includegraphics[width=\columnwidth]{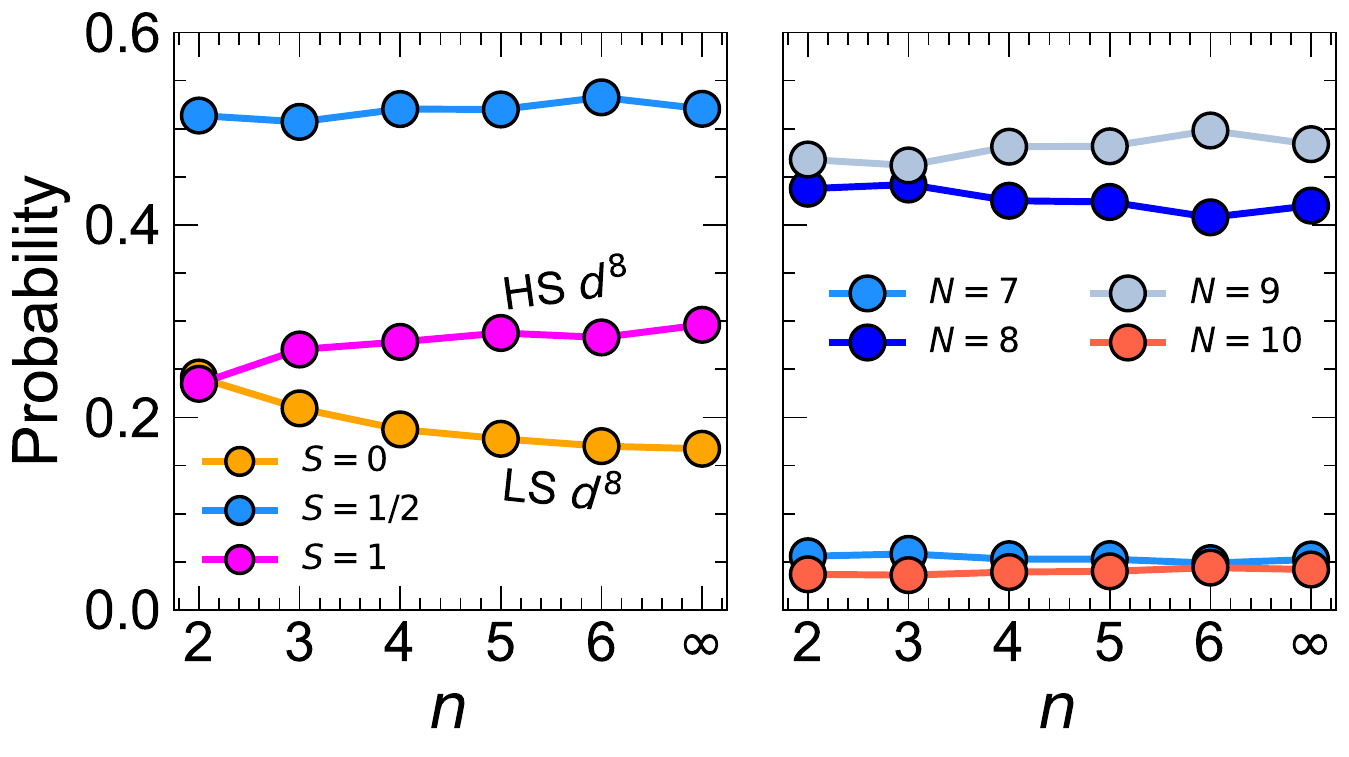}
    \caption{Most probable spin configurations (left) and occupation number (right) across the layered nickelate series $n=2-6,\infty$.}
    \label{fig:spin}
\end{figure}

\subsection{Spin and charge statistics}
The DFT+DMFT builds on an atomic picture of a solid, thus places a high importance on the atomic multiplets of the impurity problem being solved. We can gain further insight into the electronic structure of the layered nickelates by analyzing the probabilities of the atomic multiplets, which are obtained directly from the TRIQS/{\sc cthyb} solver. Figure \ref{fig:spin} shows the evolution of the three dominant spin states: $S=1/2$ ($d^{9}$), $S=1$ (high-spin $d^{8}$), and $S=0$ (low-spin $d^{8}$). Throughout the entire series, the significant charge fluctuations are between the $d^{9}$ and $d^{8}$ charge sectors with essentially no dependence on $n$ (see Fig. \ref{fig:spin}). The dominant spin configuration remains $S=1/2$, which is independent of $n$. The next two most probable configurations correspond to the low-spin (LS) or high-spin (HS) states associated with the $d^{8}$ charge sector. We find that the probability of these two spin configuration does exhibit a dependence on $n$. For the $n=2$ material, low-spin $d^{8}$ and high-spin $d^{8}$ configurations are essentially equally probable, while with increasing $n$ the high-spin $d^{8}$ configurations become dominant over the low-spin configurations \cite{Karp2020_438,petocchi2020,labollita2022}.

Table \ref{tab:occupations} summarizes how the Ni($3d$) occupations evolve across the series of layered nickelates, which have been calculated from the impurity Green's function(s). We find the total Ni($3d$) occupation remains essentially independent of $n$, in agreement with previous DFT+DMFT works for the $n=3$ and $n=\infty$ materials \cite{Karp2020_438, petocchi2020}. Within the Ni($3d)$ shell, the occupation of the $d_{x^{2}-y^{2}}$ orbital(s) monotonically decreases with decreasing $n$. This overall reduction in $d_{x^{2}-y^{2}}$ occupation is compensated by the filling of the remaining Ni($3d$) orbitals. It is important to note that these occupations should be considered on a qualitative level as they are dependent upon the projectors (i.e., projectors or Wannier functions) used to create the correlated subspace, as well as on the size of the correlated subspace \cite{Karp2021}.

The overall trend in the spin statistics can be understood from the interplay between O$(2p$) and La$(5d$) orbitals. In cuprates, the ground state electronic configuration is comprised of equal amounts of $|d^{9}\rangle$ and $|d^{10}\underbar{L}\rangle$ (where $\underbar{L}$ denotes ligand hole), which is expected of a charge-transfer material with small $p-d$ splitting. The large amount of $|d^{8}\rangle$ in the ground electronic configurations of the nickelates contrasts with that of the cuprates. In the case of the nickelates, there is a larger $p-d$ splitting (large charge-transfer energy) which increases the significance of the $d^{8}$ states, while decreasing the relevance of the $d^{10}\underbar{L}$ states. Furthermore, the larger $p-d$ splitting forces charge-transfer from the Ni sites to the La sites that function as a charge reservoir \cite{Karp2020_438}.

\begin{figure*}
    \centering
    \includegraphics[width=1.5\columnwidth]{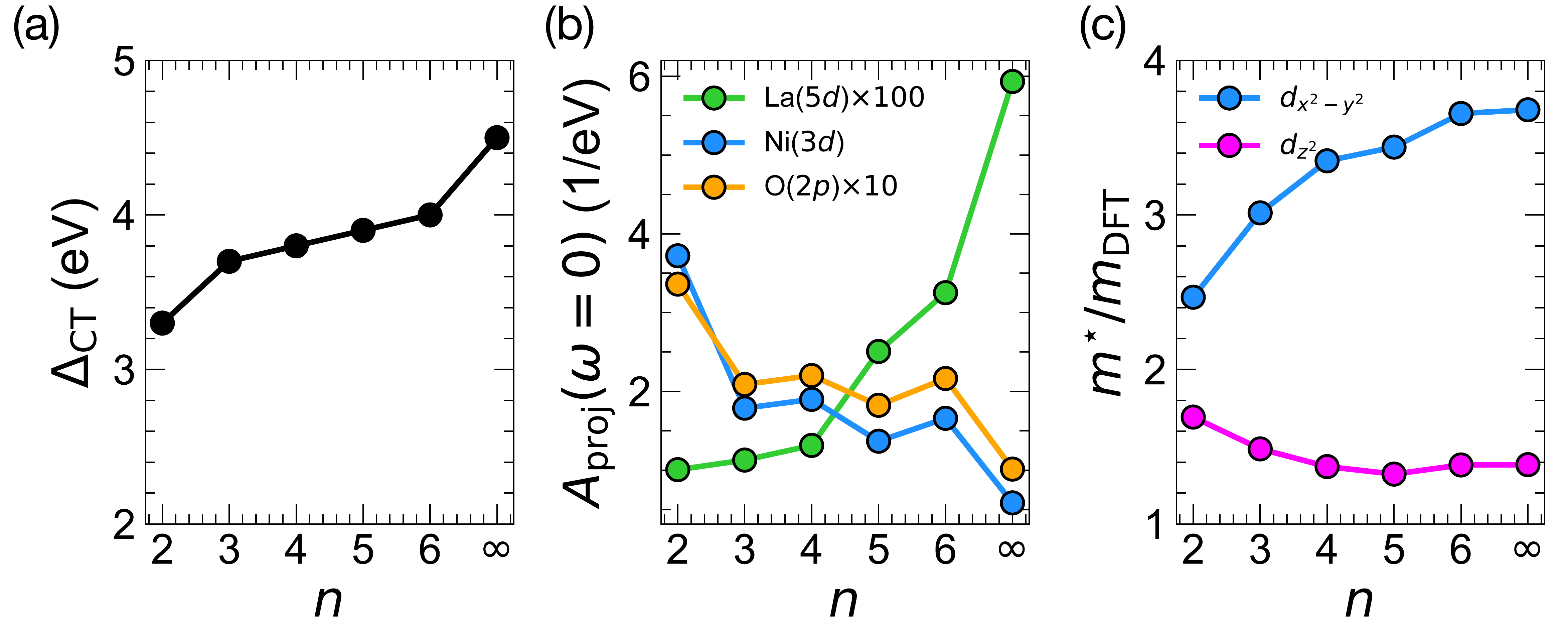}
    \caption{Key features of the electronic structure as a function of $n$: the number of NiO$_{2}$ layers. (a) Charge-transfer energy ($\Delta_{\mathrm{CT}}$), (b) $A_{\mathrm{proj}}(\omega=0)$, and (c) $m^{\star}/m_{\mathrm{DFT}}$ for the $e_{g}$ orbitals (averaged over the Ni impurities).}
    \label{fig:summary}
\end{figure*}

\footnotetext[1]{We note that the the charge-transfer energies reported in Fig. \ref{fig:summary} were calculated on the level of DFT. The charge self-consistent DFT+DMFT theory has the effect to decreases the $p-d$ splitting (the charge-transfer energy) by a small constant for each $n$. Therefore, the qualitative trend is the same at the level of both DFT and DFT+DMFT.}

\section{\label{sec:summ}Summary and Discussion}
We have employed charge self-consistent DFT+DMFT to calculate the many-body electronic structure of the layered nickelates $R_{n+1}$Ni$_{n}$O$_{2n+2}$ with $n=2-6,\infty$ within the paramagnetic state. Despite the significant change in formal valence on the Ni site, we find the general spectral properties across the series of materials are remarkably similar not only to each other, but also to the end member ($n=\infty$) which possesses a different crystal structural (while finite-layer nickelates have a spacing $R$-O fluorite block, the infinite-layer material does not). For all layered nickelates, the metallic Ni-$d_{x^{2}-y^{2}}$ states are the most strongly correlated and dominate the low-energy physics making the largest contribution to the density of states around the Fermi level. We identify key features of the electronic structure that are a function of $n$, the number of NiO$_{2}$ layers along the $c$-axis. Figure \ref{fig:summary} summarizes how these key electronic features change with $n$: we find that the number of NiO$_{2}$ layers controls the magnitude of the charge-transfer energy \cite{Note1}, the activity of the La($5d$) states around the Fermi level, and the strength of Ni($3d$) electronic correlations. Specifically, upon increasing $n$, the Ni($3d$)-O($2p$) hybridization decreases, increasing the charge-transfer energy, and the La($5d$) states shift closer to the Fermi level and begin to play an active role in the fermiology for the $n=4-6,\infty$ materials. Furthermore, we find that the electronic structure of the layered material transitions from quasi-two-dimensional to three-dimensional in the $n\rightarrow \infty$ limit. 

The layered nickelates have been classified as mixed charge-transfer-Mott-Hubbard materials \cite{Karp2020_438,dean2022} with respect to the ZSA classification scheme with our DFT+DMFT calculations further supporting this classification. In this regard, the ground state of the layered nickelate systems contains significantly more $d^{8}$ weight, which is different than a pure charge-transfer system, like the cuprates.
In spite of the increased $d^{8}$ content, these materials only contain a single strongly correlated orbital (Ni-$d_{x^{2}-y^{2}}$) and the local spectral function exhibits strong Mott-Hubbard/charge-transfer character, akin to cuprates. 
Finally, we uncover the largest difference between the finite-layer nickelates and the infinite-layer nickelates is in the dimensionality of the electronic structure. While the finite-layer nickelates have a two-dimensional-like electronic structure (mostly due to the presence of a blocking fluorite slab), the infinite-layer member has a marked three-dimensional character.  This draws the electronic structure of the finite-layer nickelates closer to that of the cuprates than the infinite-layer material, even though we caution towards the need of analyzing electronic structures at the same nominal Ni($3d$) filling to be able to do meaningful comparisons. All in all, we find similar electronic features at all $n$ suggesting the possibility that superconducting instabilities may be present in other finite-layer nickelates (beyond the $n=5$ material) if appropriate chemical doping can be achieved.

\begin{acknowledgements}
H.L and A.S.B acknowledge the support from NSF Grant No. DMR 2045826. We acknowledge the ASU Research Computing Center and the Extreme Science and Engineering Discovery Environment (XSEDE) \cite{xsede} through research allocation TG-PHY220006, which is supported by NSF grant number ACI-1548562 for HPC resources.
\end{acknowledgements}

\bibliography{ref.bib}
\end{document}